\documentclass[12pt]{iopart}
\usepackage{epsfig}

\begin{document}

\title{Coherent Waveform Consistency Test for LIGO Burst Candidates}
\author{L Cadonati }
\address{Massachusetts Institute of Technology,
LIGO Laboratory, NW17-161 Cambridge, MA 02139}
\ead{cadonati@ligo.mit.edu}

\begin{abstract}
The burst search in LIGO relies on the coincident detection 
of transient signals in multiple interferometers. 
As only minimal assumptions are made about the event waveform or duration, 
the analysis pipeline requires loose coincidence in time, frequency and 
amplitude.
Confidence in the resulting events and their waveform consistency is 
established through a time-domain coherent analysis: the \emph{r}-statistic 
test.
This paper presents a performance study of the \emph{r}-statistic test
for triple coincidence events in the second LIGO Science Run (S2),
with emphasis on  its ability to suppress the background false rate 
and its efficiency at detecting simulated bursts of different waveforms 
close to the S2 sensitivity curve.

\end{abstract} 
 
\submitto{\CQG} 
\maketitle 
 
\section{Introduction} 

The Laser Interferometer Gravitational-wave Observatory (LIGO) consists
of three detectors: H1 and H2, co-located in Hanford, WA 
and  L1, located in Livingston, LA.
The simultaneous availability of interferometric data from detectors 
with similar sensitivity and orientation allows a coherent coincidence 
analysis to be implemented in the search for bursts of gravitational waves.

In the LIGO burst analysis pipeline\cite{S1paper,gwdaw02,amaldi03}, 
candidate events are identified as excesses of power or amplitude 
in the data stream of each interferometer by a suite of search algorithms,  
referred to as \emph{Event Trigger Generators} (ETG): 
BlockNormal\cite{block1},
Excess Power\cite{Anderson01}, 
TFClusters\cite{Sylvestre02}, 
and WaveBurst\cite{waveb1,waveb2}.
The ETG tuning is tailored to maximize the detection efficiency for a variety 
of waveforms (narrow-band, broad-band and astrophysically motivated), 
with a single interferometer false rate of the order of 1~Hz.
This relatively large trigger rate  is suppressed by the multi-interferometer 
analysis, which currently only requires that events be coincident in time and 
in frequency.
The coincidence parameters (time window and frequency tolerance) are tuned 
according to the principle that the coincident detection efficiency  should 
equal the product of efficiencies in the individual interferometers.
Coincidence criteria should be loose enough not to further reduce the detection 
efficiency, within the limitations imposed by the false alarm rate.
The coincidence analysis eventually outputs triggers (start time, duration) 
when excesses of power or amplitude have been detected simultaneously 
in all interferometers.  
The first step toward validation of such events is a comparison of the 
waveforms as they appear in each detector.
 
This paper describes a test that exploits cross-correlation
between pairs of interferometers and combines them
into a multi-interferometer correlation confidence.
The test is a powerful tool for the suppression of accidental 
coincidences without reducing the detection efficiency of the pipeline.
Its performance has been tested on a 10\% portion of data from the LIGO 
second science run (S2).
  
\section{The r-statistic Cross Correlation Test}

\begin{figure} 
\begin{center} 
\epsfxsize=0.7\linewidth
\epsfbox{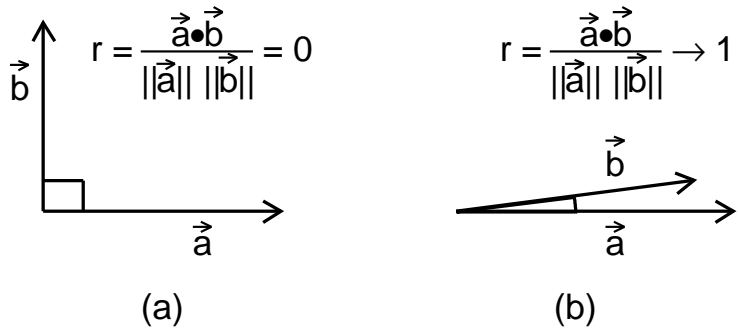} 
\end{center} 
\caption{\label{f:vector} 
Graphical representation of the $r-$statistic as cosine of 
the angle between vectors in an N-dimensional space.
(a) The two sequences are uncorrelated: the vectors are orthogonal and $r=0$. 
(b) The two sequences are correlated: as the coherent component dominates
over the incoherent noise, $r\rightarrow 1$.} 
\end{figure} 

\subsection{r-statistic}

The fundamental building block for the waveform consistency test 
is the $r$-statistic, or the linear correlation 
coefficient of two sequences $\{ x_i \}$ and  $\{ y_i \}$ :
\begin{equation}
r = \frac {\sum_i (x_i - \bar{x})(y_{i} - \bar{y})} 
{\sqrt{\sum_i (x_i-\bar{x})^2}\sqrt{\sum_i (y_{i}-\bar{y})^2}}.
\end{equation}
This quantity only assumes values between -1 (fully anti-correlated sequences), 
0 (un-correlated sequences) and +1 (fully correlated sequences). 
More generally, if the two sequences are uncorrelated, we expect 
the $r$-statistic to follow a normal distribution, with zero mean 
and $\sigma=1/\sqrt{N}$, where $N$ is the number of data points used to 
compute $r$.
A coherent component in the two sequences will cause $r$ to deviate from
the normal distribution.
If we think of data sequences as vectors in an $N-$dimensional space, 
the $r-$statistic can be seen as an estimator for the cosine of the angle 
between the two vectors (see Figure~\ref{f:vector}). 
As a normalized statistic, it is not sensitive to the relative amplitude 
of the two vectors; the advantage is robustness against
 fluctuations of detector response and noise floor. 

The number of points $N$ or, alternatively, the \emph{integration window} 
$\tau$, is the most important parameter in the construction of the 
$r-$statistic.
Its optimal value depends in general on the signal.
If $\tau$ is too large, the signal is ``washed out'' in the computation
of $r$; if it is too small, statistical considerations on the distribution of 
$r$ lose validity. 
Simulation studies show that a set of three integration times 
(20, 50 and 100 ms) is suitable for most short signals of interest
to the LIGO burst search.
A more detailed (and computationally intense) scan of integration windows
can be implemented in a targeted analysis as it was done for the LIGO 
externally triggered search described in~\cite{exttrig1,exttrig2}, 
which is also based on cross-correlation.

\subsection{Data Conditioning}
The $r$-statistic test is especially effective when all coherent lines 
and known spectral features are removed from the raw strain data;
for this reason, data conditioning plays an important role in the test. 

Raw data from each interferometer is band-passed and decimated, in order to 
suppress the contribution of seismic noise and instrumental artifacts 
at low and high frequencies and restrict the coherent analysis to 
the most sensitive frequency band in the LIGO interferometers.
For the performance studies reported in this paper, the band of interest
was 100-2048~Hz.

Next, the data is effectively whitened by a linear error predictor filter 
trained on a 10 sec period before the event start time.
This filter, described in more detail in ~\cite{lpef,rstat1}, removes 
predictable content, such as lines and the spectral shape, and emphasizes 
transients. 

 \begin{figure} 
 \begin{center} 
  \epsfxsize=\linewidth\epsffile{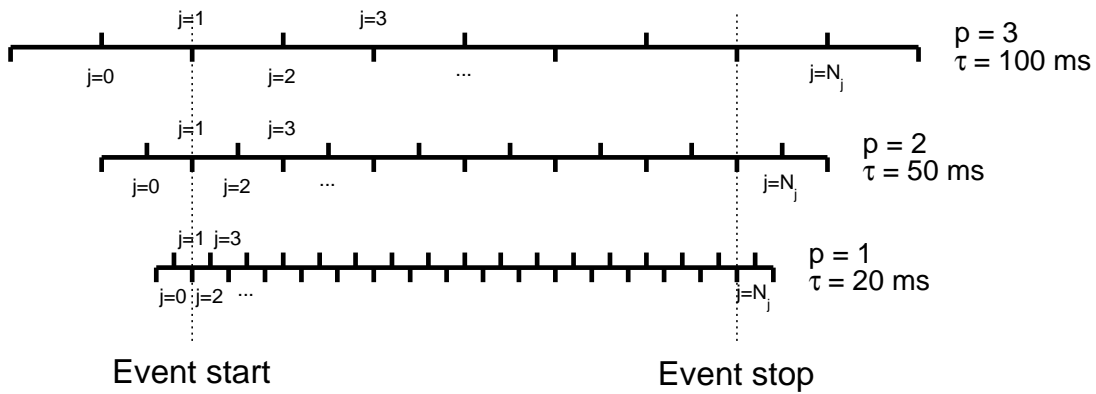} 
 \end{center} 
 \caption{\label{f:rscan} Schematic representation of how an event trigger is 
 scanned in the $r$-statistic test: for each value of the integration window 
 $\tau$, the trigger is partitioned in $N_j$ intervals of width $\tau$, with 
 50\% overlap.} 
 \end{figure} 

\subsection{Trigger Scan}
The next step consists of partitioning the trigger duration in intervals equal 
to the integration window, with 50\% overlap, as shown in figure~\ref{f:rscan}.
For each pair of interferometers $l,\,m$ in \{L1, H1, H2\}, a selected value 
of the integration window $p$ in \{20ms, 50ms, 100ms\} and each interval $j$, 
data is selected from the conditioned time series  of two interferometers.
One of the two sequences is time-shifted with respect to the other,
yielding a distribution of $r$ coefficients:
\begin{equation}
r^k_{plmj} = \frac 
{\sum_i (x^i_{plmj} - \bar{x}_{plmj})(y^{i+k}_{plmj} - \bar{y}^k_{plmj})} 
{\sqrt{\sum_i (x^i_{plmj}-\bar{x}_{plmj})^2}
 \sqrt{\sum_i (y^{i+k}_{plmj}-\bar{y}^k_{plmj})^2}},
\end{equation}
where the index $k$ represents the time lag between the two series, in steps 
equal to the inverse of the sampling rate, covering the whole $\pm 10\,$ms 
range to account for the light travel time between LIGO sites.
The quantity $r^k_{plmj}$  assumes values in $[-1,1]$ but, as the test is 
mostly interested in how much the correlation coefficient deviates from 0, 
only its absolute value $|r^k_{plmj}|$ is used.

A Kolmogorov-Smirnov test with 5\% significance is used to compare the  
$\{|r^k_{plmj}|\}$ distribution to the null hypothesis expectation of a normal 
distribution, with zero mean and $\sigma=1/\sqrt{N_p}$, where $N_p$ is the 
number of data samples in the $p$-th integration window.
If the two are inconsistent, the next step is to compute the one-sided 
significance and the corresponding confidence:
\begin{equation}
S^k_{plmj} = {\rm erfc} \left ( |r^k_{plmj}| \sqrt{\frac{N_p}{2}} \right ) ; 
\;\;\;\;\;\;\;\;\;\;\;\;\;\;\;
C^k_{plmj} = - \log_{10}(S^k_{plmj}) .
\end{equation}
A correlation confidence is assigned to interval $j$ for interferometers $l,m$
and integration window $p$ as the maximum confidence over all time lags:
\begin{equation}
\gamma_{plmj}= \max_k{C^k_{plmj}} .
\end{equation}
The degeneracy over interferometer pairs is solved through the arithmetical 
average of confidences from all combinations of $N_{ifo}$ interferometers:
\begin{equation}
\gamma_{pj}= \frac{1}{N_{ifo}(N_{ifo}-1)}\sum_{l\neq m}\gamma_{plmj} .
\end{equation}
Finally, $\Gamma$, the combined correlation confidence for the \emph{event},
is obtained maximizing over all time intervals and integration windows:
\begin{equation}
\Gamma = \max_{p}\left[\max_{j}\gamma_{pj}\right]
\end{equation}
The event passes the waveform consistency test if:
\begin{equation}
\Gamma > \beta ,
\end{equation}
where $\beta$ is the threshold imposed on the multi-interferometer 
correlation confidence.
For additional details on the method and its implementation, 
see~\cite{rstat1,rstat2}.

 \section{Triple Coincidence Performance Analysis in S2}

The $r$-statistic test performance has been explored, independently of previous 
portions of the burst analysis pipeline, by adding simulated waveforms to real 
interferometer noise and then passing 200 ms of data around the simulated peak 
time through the $r$-statistic test.

For convenience, we define here two quantities that will be used to 
characterize a burst signal. 
$h_{rss}$ is the square root of the total burst energy, in unit of 
strain/$\sqrt{\rm Hz}$: 
\begin{equation}\label{e:hrss}
h_{rss} = \sqrt{\int_0^\infty \left | h(t) \right | ^2 dt} =
\sqrt{\int_{-\infty}^\infty \left | \tilde{h}(f) \right | ^2 df} .
\end{equation}
This quantity can be compared directly to the sensitivity curves, as is also 
reflected in this definition of signal-to-noise ratio:
\begin{equation}\label{e:snr}
{\rm SNR} = \sqrt{2\int_0^\infty{\frac{ | \tilde{h}(f)  | ^2}{S_h(f)} \, df}} ,
\end{equation}
where $S_h(f)$ is the single-sided detector noise.
In particular, for narrow-band bursts with central frequency $f_0$, 
this ``excess-power'' definition of SNR becomes the ratio of $h_{rss}$ 
to the detector sensitivity at frequency $f_0$:
${\rm SNR} \approx {h_{rss}}/{\sqrt{S_h(f_0)} }$.
In the following, $S_h(f)$ will be the single-sided reference noise for the 
S2 run.

The LIGO burst search has adopted sine-gaussians as standard narrow-band
waveforms to test the search algorithms:
\begin{equation}
h(t)=h_{peak} \sin{(2\pi f_0(t-t_0)){\rm e}^{-(t-t_0)^2/\tau^2}}
\;\;\; ; \;\;\;
h_{rss}=h_{peak}\sqrt{\frac{Q}{4\sqrt{\pi}f_0}} ,
\end{equation}
where  $Q=\sqrt{2}\pi\tau f_0$ is the number of cycles folded under 
the gaussian envelope.
As instances of broad-band, limited-duration bursts, 
one can also consider Gaussians of the form:
\begin{equation}
h(t)=h_{peak} {\rm e}^{-(t-t_0)^2/\tau^2}
\;\;\; ; \;\;\;
h_{rss}=h_{peak}\sqrt{\tau\sqrt{\frac{\pi}{2}}} .
\end{equation}

Linearly polarized signals of both types and various amplitudes have been 
simulated on top of actual interferometer noise in a 10\% portion of the S2 run,
referred to as the \emph{playground}. 
In all cases, the same amplitude has simultaneously been injected in 
all three LIGO detectors, with no correction for antenna pattern effects. 
In other words, the results reported here are for optimal orientation;
this is useful to test the sensitivity of the algorithm to small signals, 
close to the noise floor.

\begin{table} [htb]
\begin{center}
\caption[]{\label{t:eff} $h_{rss}$ [strain/$\sqrt{\rm Hz}$] with 50\% detection
efficiency for Q=9 sine-gaussians at various frequencies and a 1.0 ms gaussian 
pulse.
The three columns correspond to different values of the $\beta$ threshold.
These values, computed for the S2 playground (10\% of the run)
are affected by $\sim10\%$ statistical and $\sim10\%$ systematic errors.}
\small
\begin{tabular}{lccc}
\hline\hline
waveform & $\beta=3$ & $\beta=4$ & $\beta=5$ \\
\hline
SG Q=9 f$_0=153\,$Hz  & $5.5\times 10^{-21}$ & $6.7\times 10^{-21}$ & $8.1\times 10^{-21}$ \\
SG Q=9 f$_0=235\,$Hz  & $2.6\times 10^{-21}$ & $3.2\times 10^{-21}$ & $3.9\times 10^{-21}$ \\
SG Q=9 f$_0=361\,$Hz  & $3.3\times 10^{-21}$ & $3.8\times 10^{-21}$ & $4.6\times 10^{-21}$ \\
SG Q=9 f$_0=554\,$Hz  & $4.5\times 10^{-21}$ & $5.4\times 10^{-21}$ & $6.3\times 10^{-21}$ \\
SG Q=9 f$_0=850\,$Hz  & $8.7\times 10^{-21}$ & $1.0\times 10^{-20}$ & $1.2\times 10^{-20}$ \\

SG Q=9 f$_0=1304\,$Hz & $1.8\times 10^{-20}$ & $2.1\times 10^{-20}$ & $2.2\times 10^{-20}$ \\
\hline
GA $\tau=1.0\,$ms     & $6.2\times 10^{-21}$ & $7.3\times 10^{-21}$ & $8.3\times 10^{-21}$ \\
\hline\hline
\end{tabular}
\end{center}
\end{table}

Table~\ref{t:eff} reports the resulting sensitivity of the triple-coincidence
$r$-statistic analysis, quoted as the $h_{rss}$ value that is detected 
with 50\% efficiency for three possible values of the $\beta$ threshold.
These values are affected by a $\sim10$\% systematic error, due to 
calibration, and a $\sim10$\% statistical error.

\begin{figure} 
\begin{center} 
$\begin{array}{cc}
\epsfxsize=0.5\linewidth\epsffile{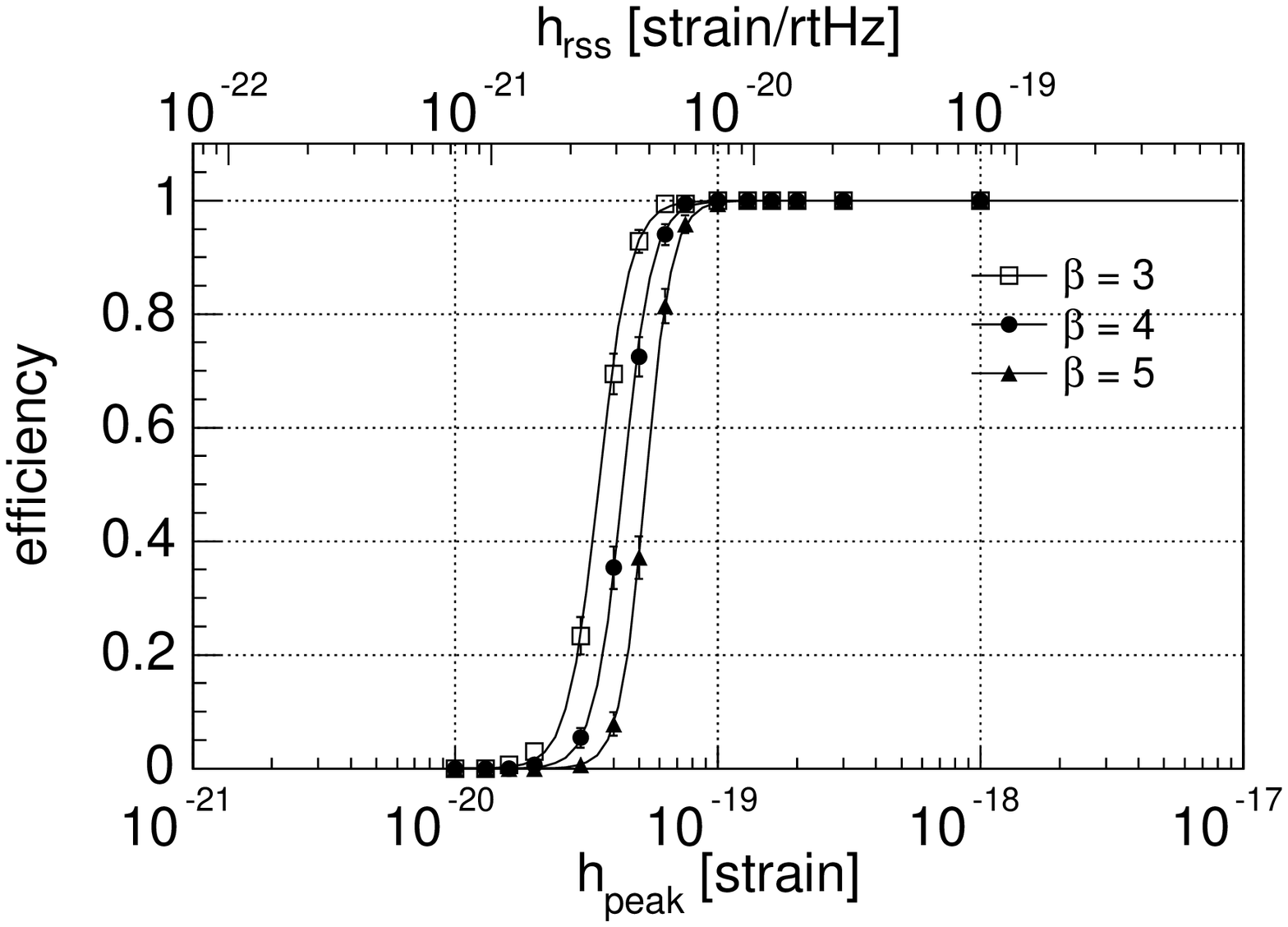} &
\epsfxsize=0.5\linewidth\epsffile{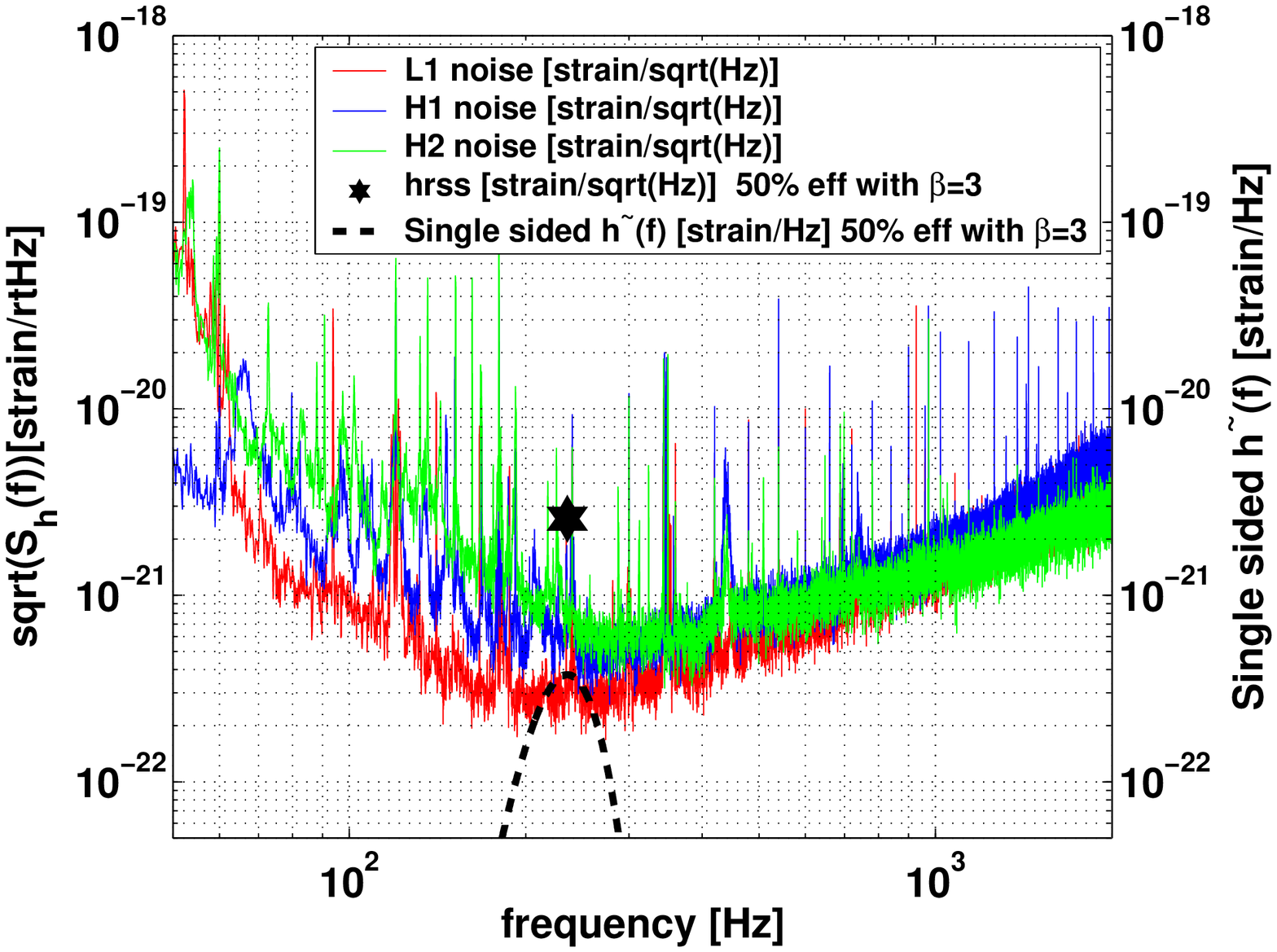} \\ 
[0.4cm]
\mbox{(a)} & \mbox{(b)}
\end{array}$
\end{center} 
\caption{\label{f:SG235} Q=9 sine-gaussian signal with $f_0=235\,$Hz.

\noindent
(a) Detection efficiency of the $r$-statistic test as function of 
the amplitude of the simulated signal. The three curves correspond 
to three values of the $\beta$ threshold.

\noindent
(b) S2 reference sensitivity curves for the three interferometers.
The star's horizontal position is the central frequency of the sine-gaussian (235~Hz);
its vertical position is the $h_{rss}$ with 50\% survival probability
if $\beta=3$ is used.
This point corresponds to $h_{rss}=2.6\times 10^{-21}/\sqrt{\mbox{Hz}}$
and SNR=9 for L1, SNR=4.5 for H1 and SNR=3.3 for H2.
Sensitivity curves and $h_{rss}$
have units of strain/$\sqrt{\rm Hz}$ (scale to the left).
The dashed curve represents the single-sided
spectrum for the corresponding waveform, in units of strain/Hz (scale to 
the right).} 
\end{figure}

\begin{figure} 
\begin{center} 
$\begin{array}{cc}
\epsfxsize=0.5\linewidth\epsffile{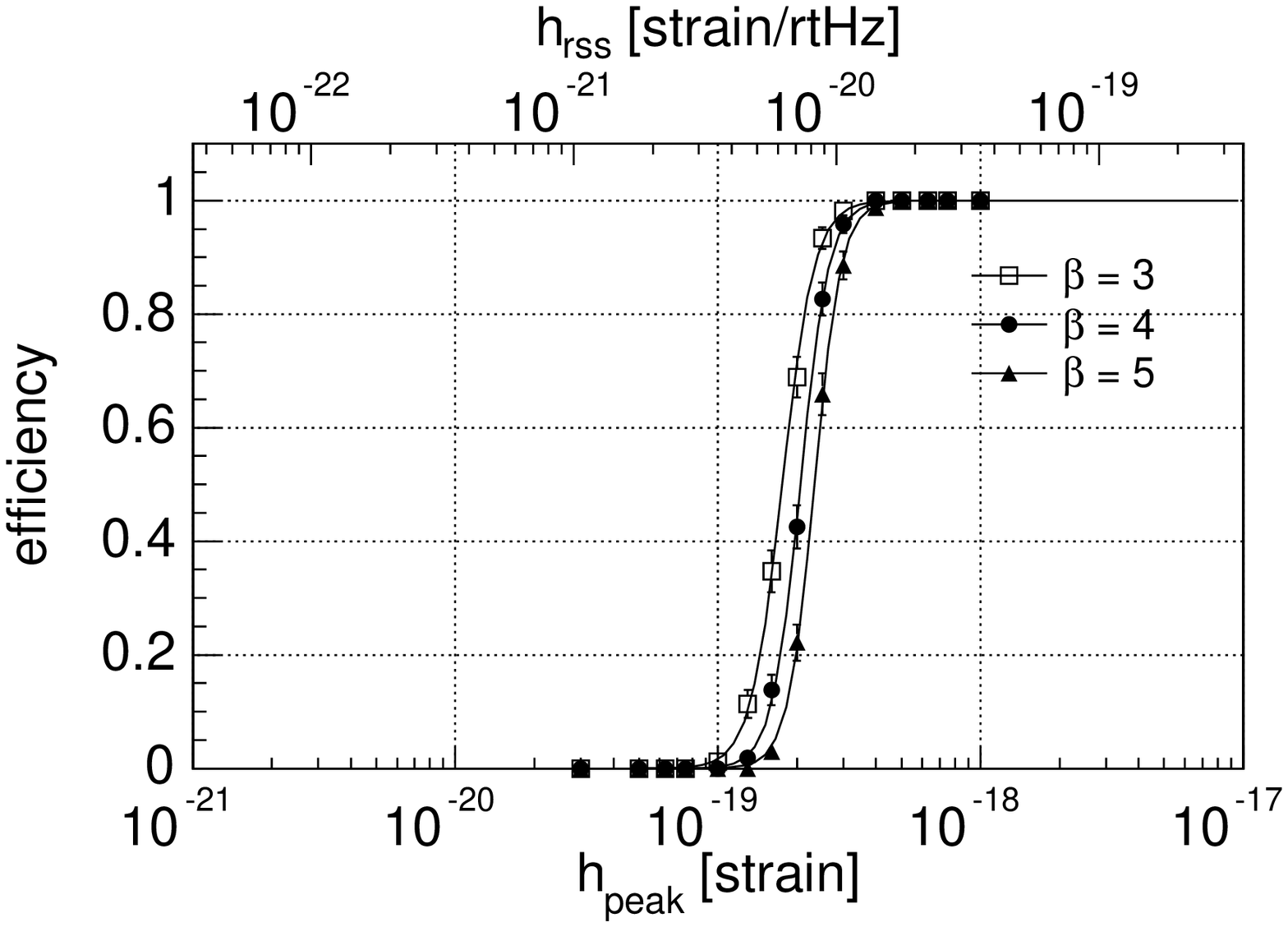} &
\epsfxsize=0.5\linewidth\epsffile{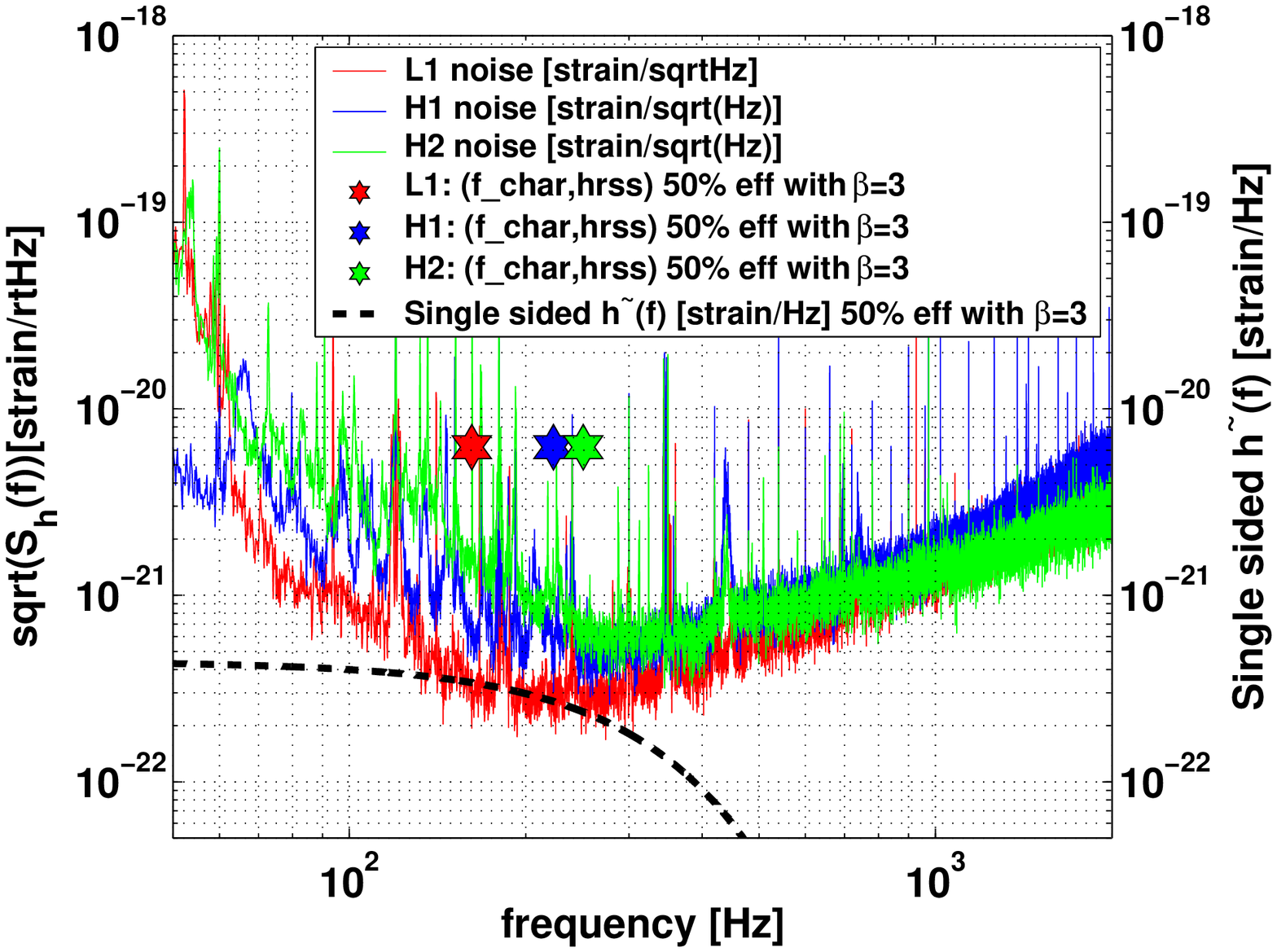} \\ 
[0.4cm]
\mbox{(1)} & \mbox{(2)}
\end{array}$
\end{center} 
\caption{\label{f:GA1ms} As Figure~\ref{f:SG235}, for Gaussian
pulses with $\tau=1.0\,$ms.

\noindent
Note that for broad band signals, the characteristic frequency,
the frequency that maximizes the SNR integrand 
$|\tilde{h}(f)|^2/S_h(f)$,
is different for the three interferometers:
$f_{char}=162\,\mbox{Hz}$ for L1,
$f_{char}=223\,\mbox{Hz}$ for H1 and
$f_{char}=251\,\mbox{Hz}$ for H2.
The 50\% detection probability, with $\beta=3$ is at 
$h_{rss}=6.2\times 10^{-21}/\sqrt{\mbox{Hz}}$
and SNR=13 for L1, SNR=6.3 for H1, SNR=4.5 for H2.} 
\end{figure}

Figures~\ref{f:SG235}~and~\ref{f:GA1ms} show, for the Q=9 235~Hz sine gaussian 
and the 1.0~ms gaussian pulses, the  efficiency curves
and the location of the 50\% point, relative to the S2 sensitivity.  
From the signal-to-noise definition in eq.~\ref{e:snr}, the 
triple-coincidence $r$-statistic test with $\beta=3$ 
results in a 50\% false dismissal, or 50\% detection efficiency, at SNR=3.3
in H2 (the least sensitive detector) for narrow-band bursts at 235~Hz or 
SNR=4.5 for 1.0~ms gaussian-like bursts.

\begin{figure} 
\begin{center} 
\epsfxsize=\linewidth\epsffile{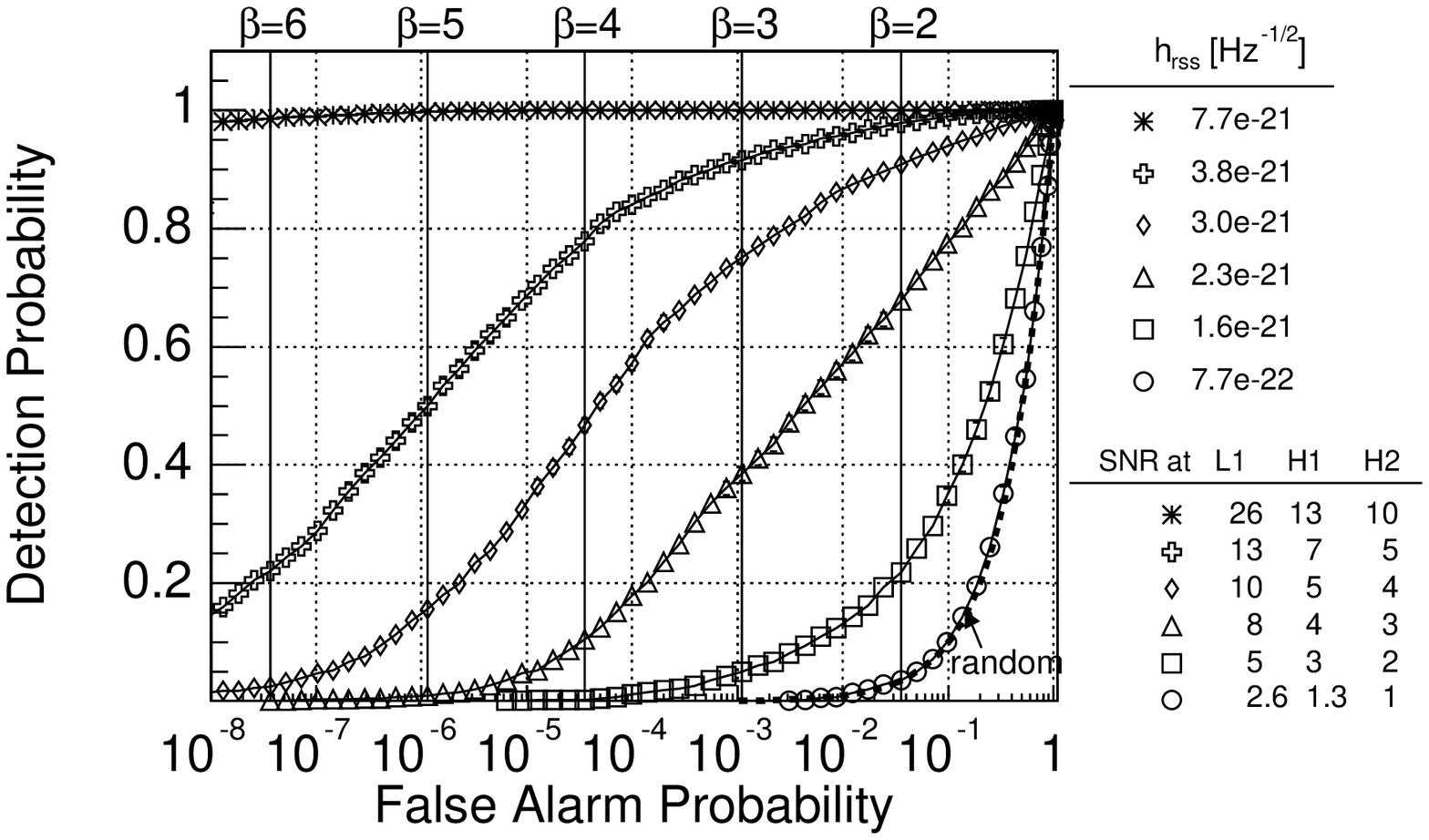} 
\end{center} 
\caption{\label{f:ROCsg235} Sine-Gaussian Q=9 $f_0=235\,$Hz: 
Receiver Operating Characteristics, or
detection probability vs false rate curves, parametrized 
with the $\beta$ threshold. Each curve corresponds to a different
signal amplitude: the top legend quotes the $h_{rss}$,
while the bottom legend shows the corresponding SNR in the three
interferometers.} 
\end{figure}

\begin{figure} 
\begin{center} 
\epsfxsize=\linewidth\epsffile{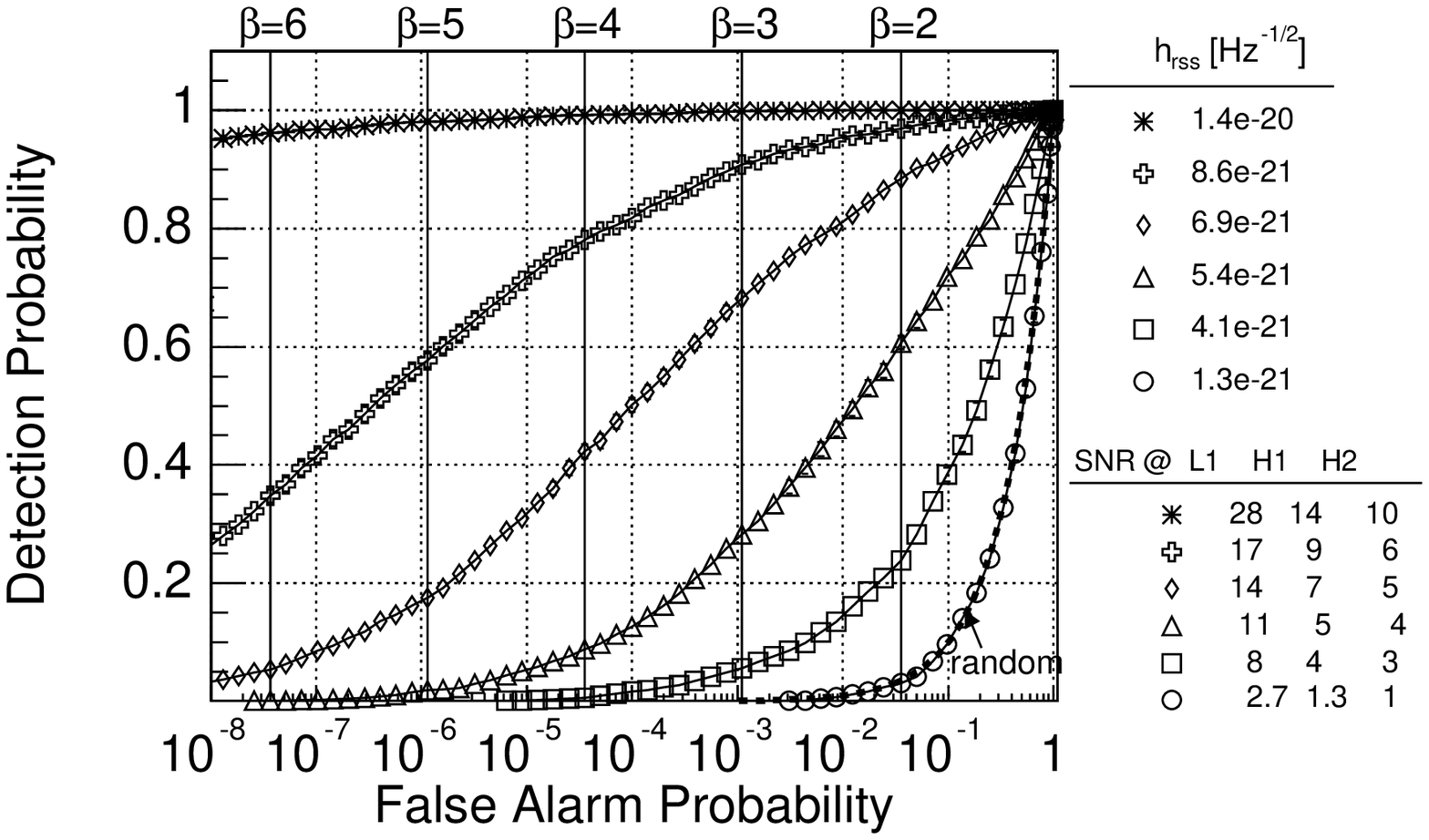} 
\end{center} 
\caption{\label{f:ROCGA1000} Gaussian $\tau=1.0\,$ms: 
Receiver Operating Characteristics, or
detection probability vs false rate curves, parametrized 
with the $\beta$ threshold.  Each curve corresponds to a different
signal amplitude: the top legend quotes the $h_{rss}$,
while the bottom legend shows the corresponding SNR in the three
interferometers.} 
\end{figure}

\begin{figure} 
\begin{center} 
\epsfxsize=0.7\linewidth\epsffile{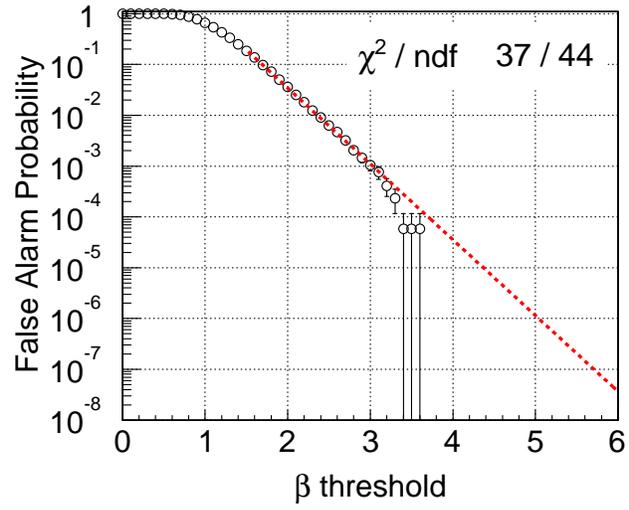}  
\end{center} 
\caption{\label{f:falserate} False rate versus $\beta$ from a sample 
of $1.7\times 10^5$ events randomly selected in the S2 playground. 
The dashed line is a fit with an exponential decay, 
applied to data points with $\beta > 1.5$.
The fit has been used to extrapolate the false rate for
$\beta > 3$ in the construction of the ROCs in 
Figures~\ref{f:ROCsg235} and~\ref{f:ROCGA1000}. } 
\end{figure}

Note that the $\beta=3(4,5)$ threshold shown here has been selected from first 
principles, as it can be tracked back to a $10^{-3}(10^{-4},10^{-5})$ false 
probability in the correlation between two interferometers on a single 
interval of duration equal to an integration window.
In order to get a more complete picture of the test efficiency for different 
values of the $\beta$ threshold, Figures~\ref{f:ROCsg235} and~\ref{f:ROCGA1000}
show, for the 235~Hz sine gaussian  and the 1.0~ms gaussian pulses, the 
detection probability versus false probability for signals between SNR=1 and  
SNR=10 relative to the least sensitive interferometer (H2).
The false probability, or the probability that an accidental event passes the
$r$-statistic test, was obtained from a sample of $1.7\times 10^5$ 200~ms 
events randomly selected in the S2 playground. 
As one can see in Figure~\ref{f:falserate}, this statistic is sufficient to
estimate the false probability only up to $\beta=3$; a fit to an exponential 
decay was used to extrapolate the false rate for the ROC curves to $\beta>3$. 
The choice of $\beta$ will ultimately be set by requirements on the false 
alarm rate in the full burst pipeline.
 
In all cases considered so far, the sensitivity of the $r$-statistic is 
comparable to or better than that of the ETGs (see for 
instance~\cite{block1,waveb1}), whose 50\% detection efficiency in triple 
coincidence typically is in the SNR=5-10 range. 
This means the $r$-statistic test with $\beta=3,4$ has a very small effect
on the detection efficiency of the burst analysis pipeline.
  
It is worth emphasizing that these false alarm probabilities have been computed 
by applying the $r$-statistic test to random times in the S2 dataset.
However, when the $r$-statistic is fully integrated in the burst pipeline
it acts on triggers that are pre-selected by the ETGs and their coincidence
analysis.
Such events share a minimal set of properties in all three interferometers:
at the very least, they are simultaneous excesses of power in overlapping 
frequency bands.
It is reasonable to expect a larger portion of these events to survive
the $r$-statistic test than what is shown above for random events, at fixed 
$\beta$. 
Nevertheless, a preliminary analysis of ETG background coincident triggers in 
S2 indicate  the $r$-statistic test can effectively suppress the accidental 
rate by $2-4$ orders of magnitude, depending on the value of $\beta$, at 
negligible cost for the detection efficiency.

\section{Conclusion}

The LIGO burst S1 analysis~\cite{S1paper} exclusively relied on event trigger 
generators and time/frequency coincidences.
The search in the second science run (S2) includes a new module of coherent
analysis: the $r$-statistic waveform consistency test.
By thresholding on $\Gamma$, the correlation confidence of coincident
events, the test can effectively suppress the burst false alarm rate
by 2-4 orders of magnitude.
Tests of the method, using simulated signals on top of real S2 noise,
yield 50\% triple coincidence detection efficiency for narrow-band and 
broad-band bursts at SNR=3-5 relative to the least sensitive detector.

\ack
The author gratefully acknowledges the support of the United States National 
Science Foundation for the construction and operation of the LIGO Laboratory 
and the Particle Physics and Astronomy Research Council of the United Kingdom, 
the Max-Planck-Society and the State of Niedersachsen/Germany for support of 
the construction and operation of the GEO600 detector.
The author also gratefully acknowledges the support of the research by these 
agencies and by the Australian Research Council, the Natural Sciences and 
Engineering Research Council of Canada, the Council of Scientific and Industrial 
Research of India, the Department of Science and Technology of India, the 
Spanish Ministerio de Ciencia y Tecnologia, the John Simon Guggenheim Foundation, 
the David and Lucile Packard Foundation, the Research Corporation, and the 
Alfred P. Sloan Foundation.

\noindent 
Particular thanks go to Shourov Chatterji, Szabolcs M\'arka
and Keith Thorne  for useful discussion and assistance in
building the infrastructure for the calculations presented in
this paper, and to the LIGO Scientific Collaboration and its Burst 
Analysis Group for the intense activity that led to the availability 
of LIGO S2 data.

\noindent
This work was supported by the US National Science Foundation 
under Cooperative Agreement No. PHY-0107417.

\end{document}